**Article Title:** AI-based data corrections for attenuation and scatter in PET and SPECT

**Authors:**
    Alan B. McMillan, PhD (corresponding author)
    Associate Professor, Department of Radiology, University of Wisconsin, Madison, WI, USA
    3252 Clinical Science Center, 600 Highland Ave, Madison, WI 53792
    amcmillan@uwhealth.org, @alan_b_mcmillan

    Tyler J. Bradshaw, PhD
    Associate Scientist, Department of Radiology, University of Wisconsin, Madison, WI, USA
    3252 Clinical Science Center, 600 Highland Ave, Madison, WI 53792
    tbradshaw@wisc.edu, @tybradshaw11

**Corresponding Author:** Alan B McMillan, PhD

**Disclosure Statement:** The Dept of Radiology at the University of Wisconsin receives support from GE Healthcare.

**Funding Acknowledgement:** Research reported in this publication was supported by the National Institute of Biomedical Imaging and Bioengineering of the National Institutes of Health under award number R01EB026708.

**Key Points:**
- Promising AI techniques are being used to improve attenuation and scatter correction for PET and SPECT imaging.
- AI-based attenuation correction can provide high quality synthetic CTs from other available images such as MRI or uncorrected PET images to improve the capability of PET and SPECT.
- AI-based scatter correction can accelerate PET reconstruction.
- AI methods can be used to estimate both attenuation and scatter simultaneously.

**Synopsis:** This article reviews the use of artificial intelligence-based methods for improving attenuation and/or scatter correction in PET and SPECT imaging.

**Abstract:** Recent developments in artificial intelligence technology have enabled new developments that can improve attenuation and scatter correction in PET and SPECT. These technologies will enable the use of accurate and quantitative imaging without the need to acquire a CT image, greatly expanding the capability of PET/MRI, PET-only, and SPECT-only scanners. The use of AI to aid in scatter correction will lead to improvements in image reconstruction speed, and improve patient throughput. This paper outlines the use of these new tools, surveys contemporary implementation, and discusses their limitations.

**Introduction**

Both PET and SPECT reconstructions require several corrections in order to yield high quality quantitative images. These corrections include: correction for random coincidences, detector dead-time, detector normalization, scattered coincidences (commonly referred to as scatter correction), and attenuation[1]. For the correction of scatter and attenuation, the tendency of electrons within the underlying tissue to induce Compton scattering due to electron interaction of PET or SPECT photons. This results in undetected annihilation events (attenuation) or the detection of anomalous coincidences (scatter). The likelihood of a photon interacting with a medium increases with both the distance traveled as well as the electron density of the underlying tissue and material through which the photons travel. Therefore, in order to accurately estimate attenuation and scatter, an additional image that provides linear attenuation coefficients of the material is needed. Historically this has necessitated the use of an external radioactive source to obtain transmission images on a PET system[2], and more recently has relied upon the use of CT to obtain this information for PET/CT and SPECT/CT. The more recent development of PET/MR has complicated this matter, where MR images are not routinely capable of identifying bone with positive contrast, making a direct conversion of MR data into a linear attenuation coefficients challenging[1].

Scattered coincidences typically account for 30-50% of the detected coincidence events in a PET scan. Scatter correction algorithms, such as the single scatter simulation method, require knowledge of the local activity concentration in addition to the attenuation coefficients of the object in order to estimate the likelihood of scattered coincidences. As the local activity concentration is only known after image reconstruction with scatter correction, iterative methods are typically required and can be time consuming to compute. While a full review of scatter corrections is not possible here, several reviews are dedicated to this problem[3–5].

In the past several years there has been a great effort to develop algorithms that utilize artificial intelligence (AI), that is, machine learning and deep learning techniques, to aid in the processing and reconstruction of medical imaging data. These techniques hold much promise in solving some of the quantitative challenges related to scatter and attenuation, particularly when additional transmission or CT imaging is unavailable or undesired in the attempt to reduce patient ionizing radiation exposure, or when patient motion degrades the effectiveness of the CT image, or for other reasons. Particularly in the development of PET/MR systems, where CT is not possible, there has been a particularly intense focus on the development of machine learning-based techniques that have demonstrated strong preliminary performance. AI-based approaches, unlike other algorithms, "learn" to perform the task at hand, which in this case is the production of a CT-like image. AI techniques use the concept of supervised learning, where examples of input images, such as an MRI scan, are paired with training labels that represent the desired output image, such as a CT. Through a complex process of backpropagation and iterative training[6,7], these algorithms are able to learn the relationships between features of the input data and the desired output. More recently, deep learning-based techniques, and in particular convolutional neural networks, have received substantial interest compared to more conventional machine learning-based techniques.

**AI-based Methods**

A block diagram of an AI-based workflow for AI-based attenuation and/or scatter correction is shown in Figure 1. Here an MR input image is paired with a CT image from the same patient to create a database of image pairs. Upon successful spatial registration of these two datasets, a deep learning network is used to train a model that synthesizes a CT-like image from only an MRI input. Once fully trained, the model is able to yield synthetic CT images that would be appropriate inputs into a conventional PET reconstruction pipeline. The performance of any given AI-based approach is going to depend on the type and structure of the underlying model that is utilized. While numerous types of models or networks could be used for a given application, the literature suggests that several types of specific model architectures are particularly effective. For example, most AI-based approaches utilize convolutional neural networks (CNNs). CNNs consist of multiple layers of synthetic neurons that learn the convolution kernels and sample weights to encode various features in both the input and desired output images[6,7]. There are many CNN-based architectures that have been developed, but two specific architectures have been the most studied for attenuation and scatter correction, shown in Figure 2. The first is the UNet, which is a nearly ubiquitous in its application to image processing in a wide range of fields[7]. This model utilizes a combination of essentially two models, one to encode the features of the input image, and another to decode the learned features into the desired output image. A key feature of the UNet is the sharing of learned features between encoding and decoding side of the network, which improves model performance. Another popular model structure is the generative adversarial network (GAN). A GAN consists of two separate models, one model as a generator, which is used to synthesize the output, and second model, a discriminator, which is used to determine the quality of the synthesized output from the generator[8].

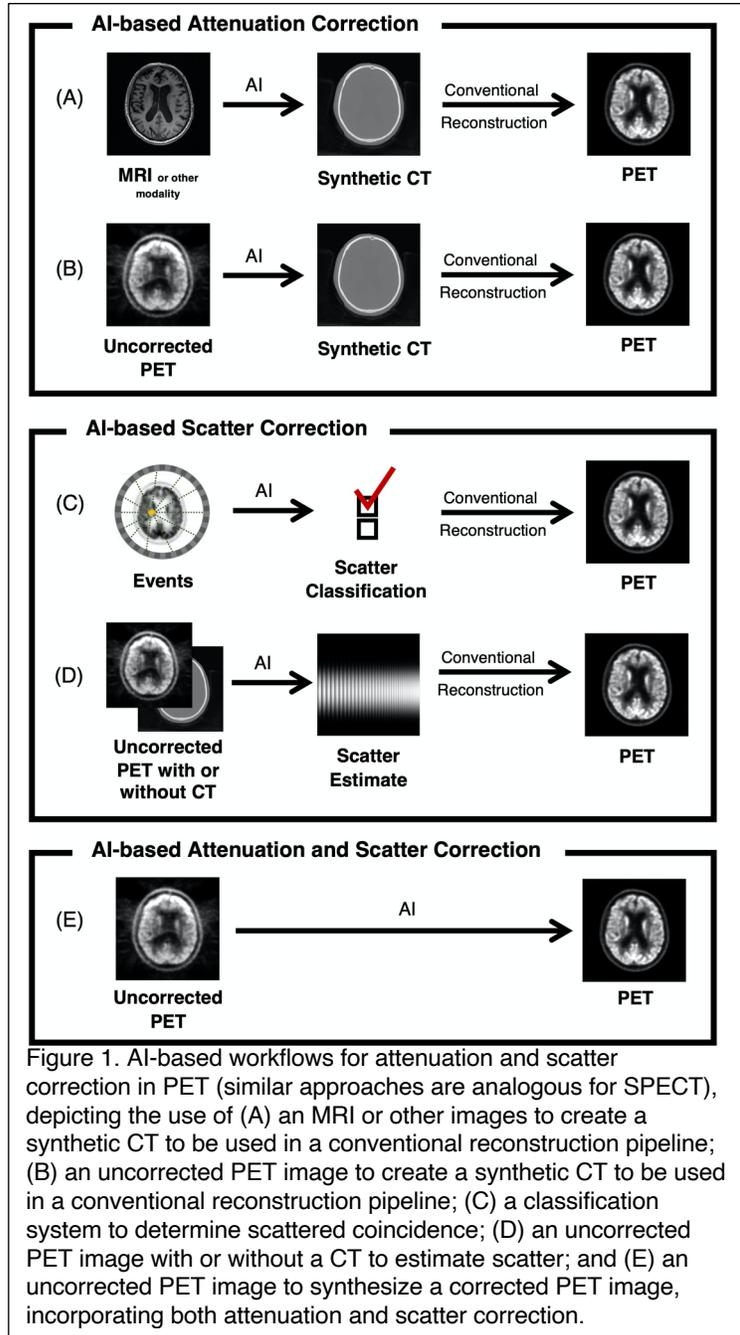

Figure 1. AI-based workflows for attenuation and scatter correction in PET (similar approaches are analogous for SPECT), depicting the use of (A) an MRI or other images to create a synthetic CT to be used in a conventional reconstruction pipeline; (B) an uncorrected PET image to create a synthetic CT to be used in a conventional reconstruction pipeline; (C) a classification system to determine scattered coincidence; (D) an uncorrected PET image with or without a CT to estimate scatter; and (E) an uncorrected PET image to synthesize a corrected PET image, incorporating both attenuation and scatter correction.

Essentially, the generator is trained such that it can create a synthesized image (e.g., a synthetic CT image) for which the discriminator cannot determine whether the generated image is real or fake. The discriminator learns from real examples and is trained to identify the fake images from the generator. The advantage of GAN-based approaches is that they can produce more realistic looking outputs. Some types of GANs do not require spatially-registered input-output pairs, which can be challenging to acquire, thus greatly reducing the burden of having paired datasets with multi-modality image registration.

*Evaluation Metrics*

The performance of a trained algorithm in synthesizing CT images is typically evaluated utilizing various metrics that compare the model's synthetic CT image to the original CT from the same patient. Common metrics to assess quality include root mean square error (RMSE), mean absolute error (MAE), structural similarity index (SSIM), peak signal to noise ratio (PSNR), and Dice coefficient (DC). Additionally, the PET or SPECT images reconstructed using the synthetic CT can be compared to the same image reconstructed using the ground truth CT. The final bias in the reconstructed image can then be quantified. Each of these metrics have their own relative strengths and weaknesses and thus the evaluation of a single model's performance should utilize multiple metrics. There exist no standard approaches for the assessment of attenuation and scatter algorithms, so investigators typically report using one or more of the aforementioned metrics. Unfortunately, this can make it difficult to compare results between different approaches.

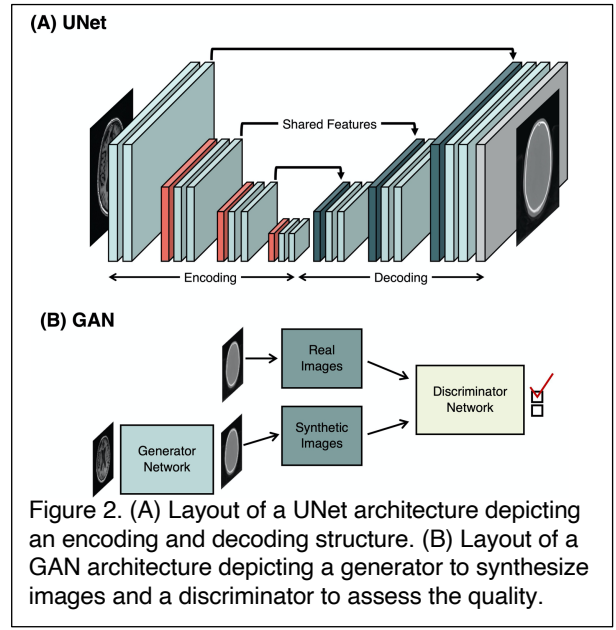

Figure 2. (A) Layout of a UNet architecture depicting an encoding and decoding structure. (B) Layout of a GAN architecture depicting a generator to synthesize images and a discriminator to assess the quality.

RMSE is a measure of the square root of the average squared difference between the synthesized CT and the original CT calculated for each voxel in the image, using the following equation:

$$RMSE(\hat{y}, y) = \sqrt{\frac{\Sigma(\hat{y}-y)^2}{n}} \quad (1)$$

, where $\hat{y}$ is the synthesized CT, $y$ is the original CT, and $n$ is the number of voxels. RMSE values are often calculated separately for soft tissue regions and bone and have units that match the original image, in this case Hounsfield units. High performing approaches typically have RMSE values of 100 HU or less.

MAE is similar to the RMSE except that it is the absolute error rather than the squared error as calculated using the following equation:

$$MAE(\hat{y}, y) = \sqrt{\frac{\sum |\hat{y}-y|}{n}} \quad (2)$$

. MAE also yields error in units that match the original image. One limitation of the MAE approach is that if errors are centered around zero, the values of MAE may be much smaller than RMSE. Both RMSE and MAE can be sensitive to the cropping of the images: if images contain large amounts of surrounding air, this can have the effect of decreasing the average error, as models can often predict surrounding air with near perfect accuracy.

SSIM is another measure of the image quality widely used throughout the computer vision literature. SSIM yields a metric on a scale that ranges between -1 to 1, where an SSIM 1 represents a perfect recovery of the original image. SSIM can be calculated using the following equations:

$$SSIM(\hat{y}, y) = \frac{(2\mu_{\hat{y}}\mu_y + c_1)(2\sigma_{\hat{y}y} + c_2)}{(\mu_{\hat{y}}^2 + \mu_y^2 + c_1)(\sigma_{\hat{y}}^2 + \sigma_y^2 + c_2)} (2)$$

$$c_1 = (K_1 L)^2$$
$$c_2 = (K_2 L)^2$$

, where $\mu_{\hat{y}}, \mu_y, \sigma_{\hat{y}}^2, \sigma_y^2, \sigma_{\hat{y}y}$ are the average, variance, and covariance of $\hat{y}$ and $y$, respectively. $K_1$ and $K_2$ are constants equal to 0.01 and 0.03. $L$ is equal to the dynamic range (i.e., the range from the smallest to largest pixel value) in the image.

PSNR is a measure of relative power of the relative difference or noise in an image compared to its reference, and can be calculated using the following equation:

$$PSNR = 10\log_{10}(L^2 / \text{MSE}) \quad (3)$$

DC is measure of overlap between discrete regions of a segmented image, where a coefficient of 1 means perfect overlap. This metric is frequently used to determine the ability of AI approaches to estimate specific regions such as air, soft tissue, and bone. DC can be calculated using the following equation:

$$\text{Dice Coefficient} = (2*TP)/(2*TP + FN + FP) \quad (4)$$

, where TP is the number of true positives, FN is the number of false negatives, and FP is the number of false positives in a segmented image relative to the reference. DC is often used to assess the quality of bone estimate in brain attenuation correction approaches. For studies in the brain, DC of 0.8 or greater in bone typically identifies a very good performing synthetic CT approach.

The end goal of CT synthesis is to use the CT for attenuation and scatter correction and create nuclear medicine images with minimum bias. Thus reconstruction bias is an important metric that compares the end result of a given attenuation or scatter correction approach to the conventional approach. Bias is typically represented as a percentage of error within a given region or over the entire image. Reconstruction image is typically calculated as follows:

$$Bias(\hat{y}, y) = \frac{y - \hat{y}}{y} \times 100\% \qquad (5)$$

Reconstruction bias averaged over the entire image can hide errors or biases occurring in small regions of high importance, such as tumors. These regions should be quantified and compared.

**AI-Based Attenuation Correction**

For AI-based attenuation correction methods, the AI algorithm synthesizes a CT-like image that can then be used in place of a CT within a conventional reconstruction pipeline. The input for the algorithm is typically an MR image (as used for simultaneous PET/MR imaging), as shown in Figure 1A, or the underlying uncorrected PET or SPECT image itself, as shown in Figure 1B. The goal of the AI algorithm is not necessarily to create a diagnostic image, but one that is sufficiently realistic image and minimizes the reconstruction bias in the final image. One shortcoming of most published AI approaches is that the models are typically trained to be tuned to identifying one anatomical region, such as the brain, chest, or pelvis and are generally not applicable to the whole body. While this does not necessarily limit these algorithms for their intended use, it does require that separate approaches be developed for each individual body part. As the feasibility of anatomic-specific algorithms for attenuation correction has already been well established, future approaches should focus on whole-body corrections, such that a uniform approach can be used regardless of the anatomy being scanned.

Several recent reviews have compared the performance of the increasing number of AI-based approaches proposed for synthetic CT generation[9–11]. In Table 1, we highlight several approaches evaluated in different body regions including the brain, chest, and pelvis for both PET and SPECT. Due to the rapid pace of innovation in this area, these results are likely to be soon joined by new and better performing approaches. However, these results demonstrate that high performing approaches already exist that appear to provide sufficient quantitative results without the need for the acquisition of a CT. Note that approaches that include a joint estimation of attenuation and scatter (Figure 1E) are listed in Table 3.

| Table 1. AI-based approaches for attenuation correction | | | | |
|---|---|---|---|---|
| Application | Approach | Advance | Quantitative Comparison vs. Acquired CT | Reference |
| Brain perfusion SPECT | CNN | Enable correction for SPECT only scanners | PSNR 62.2 dB, SSIM 0.9995 | Sakaguchi et al., 2021[12] |
| Brain PET (FDG) | CNN, UNet | Enable correction for PET only scanners | Dice coefficients of 0.80±0.02 for air, 0.94±0.01 for soft tissue, and 0.75±0.03 for bone, MAE of 111±16 HU, PET reconstruction bias <1% | Liu et al., 2018[13] |
| Brain PET/MR (FDG) | CNN | Synthesize CT from T1 MRI for PET/MRI | Dice coefficient of 0.936 ± 0.011 for soft tissue, and 0.803 ± 0.021 for bone, PET reconstruction bias −0.7% ± 1.1% | Liu et al., 2017[14] |
| Brain PET/MR (FDG) | CNN, GAN | Synthesize CT from T1 MRI for PET/MRI | Dice coefficient of 0.77 ± 0.07 for bone, PET reconstruction bias <4% | Arabi et al., 2019[15] |
| Brain PET/MR (FDG) | CNN, GAN | Synthesize CT from T1 MRI for PET/MRI | Dice coefficient of 0.74 ± 0.05 for bone, PET reconstruction bias <0.025 SUV | Gong et al., 2020[16] |
| Brain PET/MR (FDG) | CNN, UNet | Synthesize CT from T1 MRI for PET/MRI | Dice coefficient of 0.81 ± 0.03 for bone, PET RMSE 253.5, PET reconstruction bias <1% | Blanc-Durand et al., 2019[17] |
| Brain PET/MR (FDG) | CNN, UNet | Synthesize CT from T1 and ZTE MRI for PET/MRI | Dice coefficient of 0.86 ± 0.03 for bone, PET reconstruction bias <0.02 SUV | Gong et al., 2018[18] |
| Brain PET/MR (FDG) | CNN, UNet | Synthesize CT from UTE MRI for PET/MRI | Dice coefficient of 0.96 ± 0.006 for soft tissue, 0.88 ± 0.01 for bone, PET reconstruction bias <1% | Jang et al., 2018[19] |
| Brain PET/MR (FET) | CNN, UNet | Synthesize CT from UTE MRI for PET/MRI | PET reconstruction bias <1% | Ladefoged et al., 2019[20] |
| Myocardial perfusion SPECT | CNN | Enable correction for SPECT only scanners | MAE 3.60%±0.85% in whole body, 1.33%±3.80% in the left ventricle | Shi et al., 2020[21] |
| Myocardial perfusion SPECT | CNN | Enable correction for SPECT only scanners | Mean segmental errors 0.49 ± 4.35%, mean absolute segmental errors 3.31 ± 2.87% | Yang et al., 2021[22] |
| Myocardial perfusion SPECT | GAN | Enable correction for SPECT only scanners | RMSE 0.141±0.0768, PSNR 36.382±3.74 dB, SSIM 0.995±0.0043 | Torkaman et al., 2021[23] |
| Pelvis PET/MR (68Ga-PSMA) | CNN, GAN | Synthesize CT from T1 MRI for PET/MRI | PET reconstruction bias <2.2% | Pozaruk et al., 2021[24] |
| Pelvis PET/MR (FDG) | CNN | Synthesize CT from T1 and T2 MRI for PET/MRI | Dice coefficient of 0.98 ± 0.01 for soft tissue, 0.79 ± 0.03 for cortical bone, PET reconstruction bias 4.9% | Bradshaw et al., 2018[25] |
| Pelvis PET/MR (FDG) | CNN, UNet | Synthesize CT from T1 MRI for PET/MRI | PET reconstruction bias <4% | Torrado-Carvajal et al., 2019[26] |
| Whole body PET (FDG) | CNN, GAN | Enable correction for PET only scanners | PET reconstruction bias -0.8 ± 8.6% | Armanious et al., 2020[27] |

In the future, AI-based approaches may even exceed the capability of existing CT approaches, for example in the case of motion, where it might not be feasible to acquire additional CT imaging due to physiological and/or bulk patient movement. It could also improve dynamic imaging over long periods (or dual time point imaging) without needing to reacquire a CT. It appears that it may be feasible to leverage an AI-based approach to construct a motion-compensated synthetic CT or even a multi-phase CT, directly from the underlying PET or SPECT data. While there exist some potential limitations with AI-based approaches, it is likely that these approaches will gain greater clinical acceptance in the near future. Furthermore, the development of AI-based approaches for synthetic CT images to aid in attenuation correction is highly synergistic with the development of MR-only radiotherapy[28]. In this application, the valuable soft tissue contrast of MRI is useful in the delineation of tumors, yet MRI is challenged in the direct synthesis of CT-like images. Therefore, there has also been great interest in developing synthetic CT technology using AI in the radiotherapy. Developments for both improved diagnostic imaging capability in PET and SPECT as well as for radiotherapy are expected to be mutually beneficial, as the underlying techniques and AI models are highly similar.

**AI Scatter Correction**

AI based approaches to scatter correction promise to increase the overall speed of image reconstruction due to the often time-consuming nature of estimating scatter correction. Additionally, AI may allow for more accurate scatter estimation, potentially replacing single scatter simulation with an AI model that approximates multi-scatter events by using training data from Monte Carlo modeling. For example as shown in Figure 1C, one proposed approach for accelerating scatter utilizes the position and energy level of each coincidence event to predict whether an event can be categorized as originating from scatter or not[29]. Most approaches attempt to estimate the scatter profile directly from given inputs of activity as shown in Figure 1D. A summary of recent approaches to improve scatter correction (not including methods that jointly estimate scatter and attenuation) are listed in Table 2.

| Table 2. AI-based approaches for scatter correction | | | |
|---|---|---|---|
| **Application** | **Approach** | **Advance** | **Reference** |
| PET | Machine learning to classify events as true or scattered coincidences | Potential for scatter correction being performed prior to imaging reconstruction | Prats et. al, 2019[29] |
| PET | Cascade of CNNs to estimate Monte Carlo-based approach | Improved speed relative to Monte Carlo methods | Qian et. al, 2017[32] |
| PET/MR | CNN to estimate single-scatter simulation (SSS) | Significant increase in reconstruction speed | Berker et. al, 2018[30] |
| SPECT | CNN to improve filtered backprojection images to | >200x faster than Monte Carlo method | Dietze et. al, 2019[33] |

| | approximate Monte Carlo-based approach | | |
|---|---|---|---|
| SPECT, Y-90 bremsstrahlung imaging | CNN to estimate scatter projections | >120x faster than Monte Carlo method | Xiang et. al, 2020[31] |

Given that attenuation information is necessary to perform scatter estimation, and both are unknown without a CT image, many recent approaches have attempted to utilize AI approaches to determine *both* attenuation and scatter, particularly using only uncorrected PET and SPECT images as inputs. The example workflow is shown in Figure 1E and a summary of these methods are listed in Table 3.

| Table 3. AI-based approaches for simultaneous attenuation and scatter correction | | | |
|---|---|---|---|
| Application | Approach | Quantitative Comparison vs. Acquired CT | Reference |
| Brain PET (FDG, FDOPA, Flortaucipir, Flutemetamol) | CNN | PET reconstruction bias <9% | Arabi et al., 2020[34] |
| Brain PET (FDG) | CNN, UNet | PET reconstruction bias -4.2% ± 4.3% | Yang et al., 2019[35] |
| Whole body PET (68Ga-PSMA) | CNN | MAE 0.91 ± 0.29 SUV, SSIM 0.973 ± 0.034, PSNR 48.17 ± 2.96 dB, PET reconstruction bias -2.46% ± 10.10% | Mostafapour et al., 2021[36] |
| Whole body PET (FDG) | CNN | PET reconstruction bias -1.72 ± 4.22% | Shiri et al., 2020[37] |
| Whole body PET (FDG) | CNN | NRMSE 0.21 ± 0.05 SUV, SSIM 0.98 ± 0.01, PSNR 36.3 ± 3.0 dB | Yang et al., 2020[38] |
| Whole body PET (FDG) | CNN, GAN | MAE of < 110 HU, PET reconstruction bias <1%, PSNR >42 dB | Dong et al., 2019[15] |

**Challenges and Opportunities for AI Approaches**

The implementation of AI is not without challenges. There are numerous general issues that must be addressed with regards to the application of AI in medical imaging, and many of those issues affect the techniques described herein. Generalization, or the ability of a developed model to applied to a wide range on input data, and bias, or the tendency of the developed models to favor only certain situations, are important challenges in the field of AI[39]. For the applications of AI in attenuation and scatter correction, these biases affect the generalizability of a model and are thus critical to understand and address. While bias can be unintentional, it can have significant effects on the performance of models in the field compared to the tightly constrained environment of a research laboratory.

Areas of bias relevant to the techniques described herein are related to the application, population, and equipment used to develop a given AI model. If the model is applied to datasets that differ greatly in any of these categories, then the model may give spurious outputs that fall outside of the reported performance specifications. An example of application bias would be the use of an AI model for CT synthesis designed for brain imaging being applied to the pelvis. If the model was developed using only brain imaging data, it cannot be expected to perform well in other body regions. Additionally, if the model was trained with a particular MRI sequence but applied using a different MRI sequence, performance may be degraded.

Another source of bias is related to population of which the model is trained and applied to. Patients with different or abnormal anatomy of types that was not included within the training data could cause the model to give unexpected outputs. Therefore, training datasets must incorporate the type of data that is expected to be applied to the model or population specific biases (e.g., female and male, adult and pediatric, implants and amputations) will lead to unpredictable performance of model. There exist many recent demonstrations of this type of bias in AI in popular media outlets.

Finally, the specific equipment used may have a direct impact on the output of a given model. A given model may ultimately be specific to aspects of the equipment from which the images came, including, but not limited to the choice of vendor, site-specific protocols, etc. While this may be advantageous to single medical imaging equipment vendor, the ability of an AI model to be applied to multiple scanner systems is likely to be more impactful and lead to greater standardization of these powerful new methods. The key to enabling greater generalization lies in the development of multi-site and multi-vendor trials, where models are trained on data originating from different sites and equipment from different vendors. The use of federated learning techniques[40], where training is distributed to different sites, without requiring the transfer of individual images, is gaining traction as a valuable method to reduce site-specific bias and increase generalizability of developed models.

**Summary**

AI technology holds great promise to improve methods related to quantitative PET and SPECT imaging. The ability of AI-based approaches to obviate the need for an additional CT or transmission image greatly improves the capability of existing equipment and potentially reduces the cost of future systems. Integrating AI into scatter correction also leads to the ability to improve quantification as well as patient throughput by accelerating these necessary corrections. It can only be expected that these methods will continue to improve in the future and further increase capability.

**Clinical Care Points**

- Many AI-based approaches to improve attenuation correction for PET and SPECT have been demonstrated in the literature, and will likely make their way to the clinic soon.

- AI-based methods for scatter correction can improve the reconstruction time, potentially improving throughput.

- AI-based approaches can be subject bias based on the data used to train the model. Care should be taken when these methods are applied to different populations and patients with abnormal anatomy for which the AI model has not been trained to utilize.